\documentclass[twocolumn,prl,superscriptaddress,showpacs,preprintnumbers]{revtex4}

\usepackage{amsmath,dcolumn,graphicx}

\begin{document}


\title{
Self-contained Kondo effect in single molecules}

\author{C. H. Booth} 
\affiliation{Chemical Sciences Division, Lawrence
Berkeley National Laboratory, Berkeley, California 94720, USA}

\author{M. D. Walter}
\affiliation{Chemical Sciences Division, Lawrence
Berkeley National Laboratory, Berkeley, California 94720, USA}

\author{M. Daniel} 
\affiliation{Chemical Sciences Division, Lawrence
Berkeley National Laboratory, Berkeley, California 94720, USA}

\author{W. W. Lukens} 
\affiliation{Chemical Sciences Division, Lawrence
Berkeley National Laboratory, Berkeley, California 94720, USA}

\author{R. A. Andersen} 
\affiliation{Chemical Sciences Division, Lawrence
Berkeley National Laboratory, Berkeley, California 94720, USA}
\affiliation{Department of Chemistry, University of California, 
Berkeley, California 94720, USA}

\date{Phys. Rev. Lett., in press as of Oct. 28, 2005}

\preprint{LBNL-57492}

\begin{abstract}
Kondo coupling of $f$ and conduction electrons is a common feature of
$f$-electron intermetallics. Similar effects should occur in carbon ring systems
(metallocenes).  Evidence for Kondo coupling in Ce(C$_8$H$_8$)$_2$ (cerocene) 
and the ytterbocene Cp*$_2$Yb(bipy) is reported from magnetic susceptibility 
and $L_\textrm{III}$-edge x-ray absorption spectroscopy. These well-defined 
systems provide a new way to study the Kondo effect on the nanoscale, should 
generate insight into the Anderson Lattice problem, and indicate the importance 
of this often-ignored contribution to bonding in organometallics.
\end{abstract}



\maketitle

\begin{figure}[b]
\includegraphics[angle=0,width=2.5in,trim=200 200 240 170,clip=]{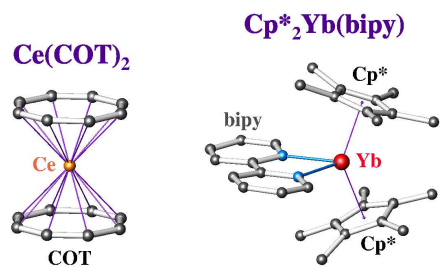}
\caption{
The crystal structure of cerocene is shown on the left, while the structure of
Cp*$_2$Yb(bipy) is shown on the right. Hydrogens atoms are not shown.
}
\label{struc}
\end{figure}

Investigations into the Kondo effect, whereby a local magnetic moment 
spin-polarizes local conduction electrons forming a magnetic singlet, has 
recently broadened from the realm of understanding heavy-fermion, mixed valent 
and other related intermetallic alloys, to the study of transport and magnetic 
properties of quantum dots \cite{Goldhaber-Gordon98,Cronenwett98}, carbon 
nanotubes \cite{Buitelaar02,Nygard00}, intermetallic 
nanoparticles \cite{Chen00}, and even single-molecule 
transistors \cite{Park02,Liang02}. Self-contained 
systems, where the conduction electrons are intrinsic rather than injected, 
are difficult to obtain experimentally, although theoretical attention has 
recently focused on magnetic impurities in $\lesssim$1 nm metallic 
nanoparticles \cite{Thimm99,Schlottmann01a,Schlottmann01b}. Similar 
interactions should occur in metal-carbon ring molecules (metallocenes) 
where the $\pi$-bonded electrons are delocalized \cite{Neumann89}. 
Here, we describe experimental evidence 
that the Kondo effect occurs in single molecules (Fig. \ref{struc})
of cerocene [Ce(COT)$_2$, where
COT = cyclooctatetraene = C$_8$H$_8$] and the ytterbocene Cp*$_2$Yb(bipy) 
[Cp* = pentamethylcyclopentadienyl = C$_5$Me$_5$, 
bipy = 2,2$^\prime$-bipyridyl = (NC$_5$H$_4$)$_2$] from 
both temperature-dependent magnetic susceptibility $\chi(T)$ and $f$-occupancy measurements using 
rare-earth $L_\textrm{III}$-edge x-ray absorption near-edge spectroscopy 
(XANES).  
These results not only provide a new arena for studying the Kondo 
effect on the nanoscale, but also indicate the importance of this often-ignored 
contribution to bonding in organometallics and provide insight into the more 
general Anderson Lattice problem.

The main difference between the Kondo effect in a bulk system and a 
nanoparticle is due to quantum confinement, where the system is small enough 
that the conduction band is no 
longer continuous, instead forming a set of discrete states with energy gap 
$\Delta$ \cite{Thimm99,Schlottmann01a,Schlottmann01b}. 
This problem is essentially a particle-in-a-box coupling to a magnetic 
impurity, and so has been dubbed a ``Kondo box" \cite{Thimm99}.
In a conventional metallic lanthanide Kondo system with a continuous density of 
states at the Fermi level, calculations show that $\chi(T)$ approaches a 
constant $\chi_0$ as $T\rightarrow 0$, and for magnetic quantum numbers $J>1$, 
it goes 
through a maximum near $T\approx\frac{1}{4}T_\textrm{K}$ \cite{Rajan83}.  In a 
nanoparticle for $T>\Delta$, the system will behave similarly to the bulk, 
continuously filling the high-$T$ triplet state from the low-$T$ singlet state, 
that is, changing from a reduced $\chi(T)$ at low $T$ to a 
Curie-Weiss $\chi(T)$ at high $T$. Both nanoparticle and bulk systems
are in a multiconfigurational, quantum mechanically mixed ground 
state of the $n_f$=0 and 1 states, with the $f$-electron (for Ce) or
$f$-hole (for Yb) occupancy $n_f$ in the range of $\sim$0.7-1.0.  
For $T_\textrm{K}\gtrsim 1000$~K,
$n_f$ should change little with $T$ \cite{Bickers87}.

The situation changes for $\chi(T)$ at $T<\Delta$, although exactly how 
probably depends on the details of the system in question. One possibility is 
that $\chi(T) \rightarrow 0$ as $T\rightarrow0$ and is exponentially
activated as $T$ approaches $\Delta$ \cite{Schlottmann01a,Schlottmann01b}, 
similarly to a Kondo insulator \cite{Riseborough00}. Implicit in this 
calculation is that the Land\'{e} $g$-values of the $f$-ion and conduction 
electrons are the same.  If one instead allows different $g$'s 
(eg., $g$=6/7 for Ce and 2 for conjugated $\pi$ electrons), one should obtain a 
$T$-independent, van Vleck paramagnetic state as 
$T\rightarrow0$~\cite{Schlottmann_priv05}.
Another possibility \cite{Thimm99} is that the Kondo resonance survives even up 
to $\Delta/T_\textrm{K} \sim 5$, and therefore a continuous density of states 
exists. If the resonance width is proportional to 
$T_\textrm{K}$ as in the bulk model, one expects a partial filling of the 
states above $\Delta$, and $\chi(T)\rightarrow \chi_0$ at low 
$T$. 
Thus, even in a nanoparticle, one may obtain qualitatively similar behavior in 
$\chi(T)$ to a bulk system.
A molecular Kondo system should allow 
exploring these issues in a well-defined, monodispersed system that is small 
enough to create large energy gaps.  

We begin by investigating cerocene and comparing these results to previous 
theoretical predictions \cite{Neumann89,Dolg91,Dolg95}. The high 
measured $T_\textrm{K}$ prevents the observation of any $T$-dependent 
effects. Consequently, Cp*$_2$Yb(bipy) is investigated as an example of a molecule 
with a lower $T_\textrm{K}$ where clear $T$-dependent effects are 
observed. Deviations between a conventional (bulk) Kondo model and these data are 
noted, and possible explanations are discussed.

\begin{figure}[b]
\includegraphics[angle=0,width=2.7in,trim=15 15 0 15,clip=]{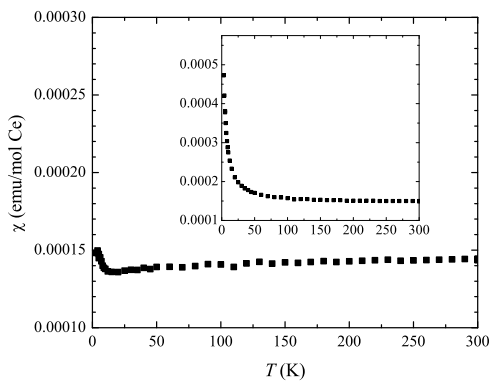}
\caption{
Cerium contribution to the magnetic susceptibility of cerocene, Ce(COT)$_2$. 
The 
diamagnetic contribution from the (COT)$_2$ has also been removed using Pascal 
constants \cite{Mulay76}.
Inset: data including a small magnetic impurity ($\sim$0.2\% of a 
$J$=5/2 impurity), removed in data from main panel. 
}
\label{cero_mag}
\end{figure}

Cerocene (Fig.~\ref{struc}) might seem like a strange place to look for Kondo 
interactions, because the required local moment would be due to partial 
occupancy of the Ce $f$-orbital. Since its initial synthesis nearly 30 
years ago \cite{Greco76}, Ce in cerocene was believed to have a 
tetravalent, $f^0$ ground state. Atomic radii arguments \cite{Raymond80} and 
cerocene's apparent diamagnetism are consistent with the nominal Ce(IV) 
assignment, and gas phase photoelectron spectra \cite{Streitwieser85}
show no obvious signs of $f$-electron ionization. However, Dolg, Fulde, and 
co-authors \cite{Neumann89,Dolg91,Dolg95} later postulated that the ground 
state of cerium is close to Ce(III) $f^1$ with an admixture of Ce(IV) $f^0$ 
character. In this picture, strong correlations between the C  
$\pi$-electrons and the Ce $f$-electrons produce a singlet ground state, as in the Kondo 
interaction. Multiconfiguration interaction calculations indicate that this 
$f$(Ce)/$p$(C) coupling is very strong, leading to a mixed-valent ground state 
with an $f$-occupancy of $n_f \approx 0.80$ and a 
$T_\textrm{K} \approx 11,600$~K. The only experimental evidence supporting 
this conclusion are Ce $K$-edge XANES spectra \cite{Edelstein96} of the 
substituted cerocenes Ce($\eta^8$-1,4-(TMS)$_2$C$_8$H$_6$)$_2$ 
and Ce($\eta^8$-1,3,6-(TMS)$_3$C$_8$H$_5$)$_2$ 
indicating that the $K$-edge position is similar to that 
from various Ce(III) model compounds.

In order to enhance previous studies of the physical properties of cerocene, a better 
synthesis was developed as the literature synthesis only yields small amounts of 
product, making purification difficult. High yields were obtained by oxidizing the anion, 
[Ce(COT)$_2$]$^-$ to cerocene using ferrocenium salts or para-quinone. 

\begin{figure}[b]
\includegraphics[angle=0,width=2.7in,trim=15 15 0 15,clip=]{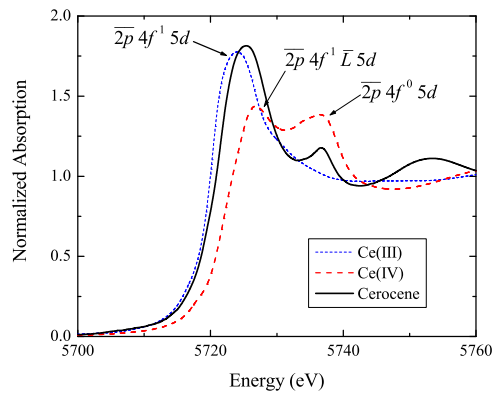}
\caption{
Ce $L_\textrm{III}$ XANES of cerocene, and  Ce(III)  
and Ce(IV) references (see text). 
Final state configurations for the references, as determined in the 
literature, are indicated.
}
\label{cero_xanes}
\end{figure}

Since the metallocenes are extremely sensitive to air, special 
sample holders for both the $\chi(T)$ and the XANES measurements are used. Susceptibility
samples were sealed into 3 mm-diameter quartz tubes, and held in place with a 
small amount of quartz wool. The total background was typically 
$\sim2\times10^{-5}$~emu (eg. $\sim$30\% of the signal at 300~K for cerocene). 
XANES holders consist of an aluminum body with machined slots. The samples 
(typically about 5 mg of ground 
powder) were mixed with boron nitride and packed into the slots. Pinhole-free 0.001" 
aluminum windows were affixed to the holder with a lead-wire seal.  This design 
allows measurements between 10-600~K while maintaining isolation from air.

Magnetic susceptibility was measured with a Quantum Design superconducting 
quantum interference device (SQUID) magnetometer. Small 
impurity contributions (so-called ``Curie tails") are removed.  For cerocene, a fit to a 
Curie-Weiss form plus a constant yields an impurity contribution with Curie constant
$C_J$=0.00167~emu$\cdot$K/mol and Weiss temperature $\Theta_\textrm{CW}=-9.1$~K, 
corresponding to $\sim$0.2\% $J$=5/2 impurity.  The diamagnetic contribution from the 
complex has also been removed within 1\% 
using Pascal corrections \cite{Mulay76}. For example, while the cerocene molecule is 
diamagnetic with $\chi(T)=-(2.7\pm0.2)\times10^{-5}$~emu/mol, the Pascal constants 
indicate that the 
(COT)$_2$ portion is $\chi(T)=-1.67\times10^{-4}$~emu/mol. The remaining cerium ion susceptibility is 
therefore positive with a similar magnitude. 

Figure \ref{cero_mag} shows the cerium contribution to $\chi(T)$ for cerocene after 
background corrections both with and without the impurity contribution. The 
cerium ion displays characteristics of $T$-independent paramagnetism (TIP) 
with $\chi(T)=(1.4\pm0.2)\times10^{-4}$~emu/mol. Estimating of 
$T_\textrm{K}$ from this result depends on whether one assumes a continuum of 
states, as discussed earlier. If the TIP is due to van Vleck paramagnetism, we 
expect $\chi_0 \sim 1/T_\textrm{K}$, since $T_\textrm{K}$ is the difference in 
energy between the singlet and triplet states.  However, we don't know the 
proportionality constant without a more detailed calculation. If we assume the 
Kondo resonance generates a continuum of states, one can use the 
Coqblin-Schriefer model to obtain 
$T_\textrm{K}=\frac{2J+1}{2\pi}\frac{C_J}{\chi_0}$ \cite{Rajan83}
where $J$=5/2, $T_\textrm{K}\approx 0.770/\chi_0$=5,500~K from these data, in 
rough agreement with the value of 11,600~K from the calculation \cite{Dolg95}. 

\begin{figure}[b]
\includegraphics[angle=0,width=2.7in,trim=15 15 0 15,clip=]{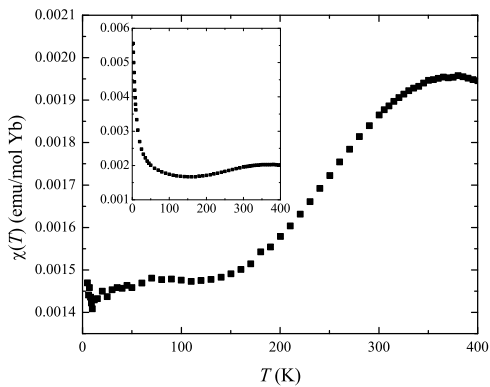}
\caption{
Ytterbium contribution to the magnetic susceptibility of Cp*$_2$Yb(bipy). 
Inset: data including impurity ($\sim$1\% of a $J$=7/2 
impurity), removed in data from main panel.
}
\label{bipy_mag}
\end{figure}

XANES data were collected on BL 11-2 at the Stanford Synchrotron Radiation 
Laboratory (SSRL) using a Rh-coated mirror to reject energies $\agt9$~keV. The double Si(220) 
monochromator crystals were detuned for further rejection. Samples were placed in a 
LHe-flow cryostat.
Figure~\ref{cero_xanes} shows the XANES data for cerocene and two reference standards: a 
Ce(III) standard, Ce[N(Si(CH$_3$)$_3$)$_2$]$_3$, and a Ce(IV) standard, 
Ce[5,7,12,14-Me$_4$-2,3:9,10-di-benzo[14]hexaenato(2-)N$_4$]$_2$ = Ce(tmtaa)$_2$. 
It is important to note that formally Ce(IV) 
systems generally are strongly mixed valent \cite{Kotani88}, as first observed 
in CeO$_2$ and later in 
other formally Ce(IV) systems \cite{Kaindl88}. The Ce(III) standard has XANES typical of the $L_\textrm{III}$ 
edge of a rare earth in a single valence configuration, displaying a sharp resonance just 
above the threshold, followed by comparatively small oscillations due to elastic 
scattering of the photoelectron by neighboring atoms. 
The final state includes the core hole 
($\overline{2p}_{3/2}$) and the excited state 5$d$, that is, 
$\overline{2p}_{3/2} f^1 5d$. Like CeO$_2$, the initial state of the Ce(IV) 
standard Ce(tmtaa)$_2$ is a superposition of states, close to 
$\frac{1}{2}|f^0>+\frac{1}{2}|f^1\overline{L}>$ , where  $\overline{L}$
indicates a ligand hole. The interaction of the $\overline{2p}_{3/2}$ core 
hole with the $f$ orbital splits the final state into 
$\overline{2p}_{3/2} f^0 5d$ and $\overline{2p}_{3/2} f^1\overline{L} 5d$ 
configurations \cite{Kotani88}. This splitting is clearly visible in the Ce $L_\textrm{III}$-edge 
spectra of cerocene and allows a more precise estimate of the cerium valence than the 
previous Ce $K$-edge work. Indeed, this spectrum unequivocally shows that while cerium 
in cerocene is much closer to Ce(III) than the Ce(IV) model compound 
in agreement with the previous measurement \cite{Edelstein96}, it also displays a pronounced feature 
indicative of an $f^0$ component and therefore is mixed valent. These 
data have been fit with a combination of an integrated pseudo-Voigt to simulate the main 
edge and other pseudo-Voigts to model the resonance features. We estimate 
$n_f=0.89\pm0.03$ from these fits. We observe no change in $n_f$ with $T$ until the 
system decomposes above $\sim$565~K. These data compare favorably with the 
value of $n_f=0.80$ obtained in the calculations \cite{Dolg95}. Moreover, the measurement is consistent 
with a strongly mixed valent system as expected from the high predicted value of
$T_\textrm{K}$ \cite{Bickers87}.

\begin{figure}[t]
\includegraphics[angle=0,width=2.7in,trim=15 15 0 15,clip=]{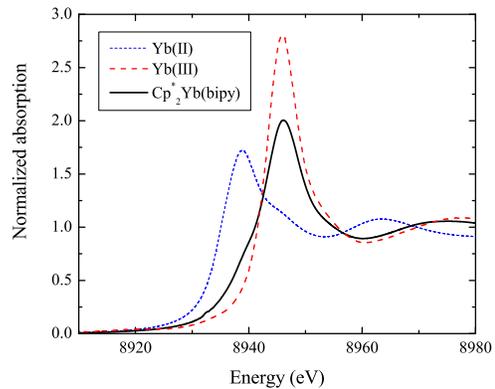}
\caption{
Yb $L_\textrm{III}$ XANES for Cp*$_2$Yb(bipy), and Yb(II) and Yb(III) 
references (see text).
}
\label{bipy_xanes}
\end{figure}

Next we study how the ground state and $T$ dependence evolve in a 
system with a lower $T_\textrm{K}$, the ytterbocene Cp*$_2$Yb(bipy) (see 
Fig.~\ref{struc}), where Yb is 
regarded as the hole analogue to Ce, and similar magnetic behavior is expected. 
In contrast to cerocene, the $f$-electrons couple with the bipy 
$\pi$-bonded electrons, since $f$-orbital overlap with $\pi$-orbitals of the canted Cp*'s
is small.

Samples of Cp*$_2$Yb(bipy), as well as Cp*$_2$Yb(OEt)$_2$ and Cp*$_2$Yb(bipy)I as Yb(II) 
and Yb(III) reference compounds, respectively, were synthesized and crystal structures 
obtained using previously discussed methods \cite{Schultz02}. 
Magnetic susceptibility data for the ytterbium contribution to Cp*$_2$Yb(bipy) 
are shown in Fig.~\ref{bipy_mag} after the Pascal corrections, with and 
without the impurity contribution removed. 
The impurity contribution in the Cp*$_2$Yb(bipy) data is slightly non-Curie-Weiss-like at low 
$T$, that is, $1/\chi(T)$ is not linear with $T$. This nonlinearity
is observed in other ytterbocenes synthesized in our lab. We overcome this difficulty by 
modeling the impurity contribution with data from the non-magnetic Cp*$_2$Yb(II)(py)$_2$ 
system, where the impurity contribution corresponds to less than $\sim$0.5\% of 
a $J$=7/2 impurity. The total impurity in the Cp*$_2$Yb(bipy) data is roughly twice as large.

These $\chi(T)$ data show significant $T$ dependence in $\chi(T)$, with clear 
evidence of both TIP at low $T$ with $\chi_0=(1.46\pm0.02)\times10^{-3}$~emu/mol
and a strong maximum $\chi_\textrm{max}$ in the susceptibility at 
$T(\chi_\textrm{max})\approx380$~K. 
Yb $L_\textrm{III}$-edge XANES data (Fig.~\ref{bipy_xanes}) are consistent with 
the $\chi(T)$ data. In this case, the $f^{14}$ configuration is non-magnetic, 
and the estimated $f$-hole occupancy 
$n_f$=0.07, 0.80, and $1.00\pm0.03$ for the Cp*$_2$Yb(II)(OEt)$_2$, 
Cp*$_2$Yb(bipy), and Cp*$_2$Yb(bipy)I, respectively. No change in these 
values has been observed from 10~K up to the $T$ at which the samples 
decompose, above $\sim400$~K.

Although there have been no detailed theoretical calculations on 
Cp*$_2$Yb(bipy), the $T$ dependence and the observed $\chi_\textrm{max}$ in 
addition to the measured mixed valence make a strong case for Kondo-like 
interactions. In this instance, assuming a continuum of states and $J$=7/2, we 
estimate $T_\textrm{K}$ from 
$T_\textrm{K}=3.27/\chi_0=2240$~K and 
$T_\textrm{K}=4.4T(\chi_\textrm{max})=1670$~K; these 
estimates are in reasonable agreement. Moreover, we measure 
$\chi_\textrm{max}/\chi_0=1.34$, compared to the $J$=7/2 
calculation of 1.22 \cite{Rajan83} , also in reasonable agreement 
especially when one considers that 
$\chi_\textrm{max}/\chi_0$ grows with increasing mixed valence
in the non-crossing approximation (NCA) \cite{Bickers87}.
If one instead assumes discrete states and a 
van Vleck TIP as $T\rightarrow 0$, the observed
$T$-dependence would be due to activation to the triplet level.
Since $T_\textit{K}$ estimates from both $\chi_0$ and 
$T(\chi_\textrm{max})$ assuming a continuum of states are similar, we tend to
favor the idea that a Kondo resonance induces bulk-like behavior, but
the discrete-states, van Vleck model cannot be ruled out from these data.

Evidence of activated behavior in $\chi(T)$ is not observed for either the 
cerocene or ytterbocene molecules, implying that within the continuum of states 
picture $\Delta$ is $\lesssim0.2$~eV, and that a lower 
$T_\textrm{K}$ is still required to observe the predicted activated behavior. 
It is important to point out that the enhanced $\chi_\textrm{max}/\chi_0$ ratio 
may be due to a size-induced suppression of $\chi_0$ as $T_\textrm{K}$ 
approaches $\Delta$, also accounting for the higher estimate of $T_\textrm{K}$ 
from $\chi_0$ compared to that from $T(\chi_\textrm{max})$. In fact, our early 
results on some related molecules show that $\chi_\textrm{max}/\chi_0$ grows as 
$n_f \rightarrow 1$, opposite to that expected from calculations in the 
NCA \cite{Bickers87}.  Another important possibility is that the number of 
electrons available for screening the $f$-hole moment is close to unity. In that
case, the screening cloud could be over or under damped and the susceptibility 
can approach the high-$T$ local moment state either more quickly or more slowly 
with $T$ than in a single impurity model \cite{Tahvildar-Zadeh97}.  This issue 
therefore directly links these materials to current topics in understanding the
Anderson Lattice.

In summary, the Kondo model has been used to understand the electronic and 
magnetic behavior of cerocene Ce(COT)$_2$ and the ytterbocene Cp*$_2$Yb(bipy). 
The experimental results for cerocene are in close agreement with theory, both 
from pseudopotential calculations \cite{Dolg95} and bulk Kondo calculations. In 
particular, $\chi(T)$ of cerocene is $T$ independent, consistent with the 
high value of $T_\textrm{K}$, and the valence is close to Ce(III) with some Ce(IV) 
character. Data on Cp*$_2$Yb(bipy) are qualitatively similar, 
except that $T_\textrm{K}$ is low enough that $\chi(T)$ is $T$-dependent,
showing TIP susceptibility below $\sim$150~K and a clear maximum at about 380~K,
consistent with the measured valence within the Kondo model. Finally, deviations
from the standard Kondo picture exist, possibly due to size effects.

\acknowledgments

We thank E. D. Bauer, D. L. Cox, M. Crommie, J. M. Lawrence, J. L. Sarrao and
P. Schlottmann for 
many useful discussions.  This work was supported by the Director, Office of Science, Office of 
Basic Energy Sciences (OBES), Chemical Sciences, Geosciences and Biosciences Division, U.S. 
Department of Energy (DOE) under Contract No. AC03-76SF00098. XANES data were collected at the 
SSRL, a national user facility operated by Stanford University of 
behalf of the DOE/OBES.

\bibliographystyle{prsty}
\bibliography{/home/hahn/chbooth/papers/bib/bibli}

\end{document}